\input{aipcheck}

\documentclass[
    ,final            
  ]
  {aipproc}

\layoutstyle{6x9}

\newcommand*{\bea}{\begin{eqnarray}}
\newcommand*{\eea}{\end{eqnarray}}
\newcommand*{\be}{\bea}
\newcommand*{\ee}{\eea}
\newcommand*{\pd}{\partial}
\newcommand*{\pdm}{\pd_{\mu}}

\newcommand*{\mn}{{\mu\nu}}

\begin{document}

\title{Propagators in Yang-Mills theory \\ for different gauges}


\author{Axel Maas}{
}

\author{Attilio Cucchieri}{
}

\author{Tereza Mendes}{
  address={Instituto de F\'\i sica de S\~ao Carlos, Universidade de S\~ao Paulo,
           Caixa Postal 369, \\ 13560-970 S\~ao Carlos, SP, Brazil}
}

\begin{abstract}
Green's functions are gauge-dependent quantities. Thus, the manifestation of confinement
in these correlation functions also depends on the gauge. Here we use lattice gauge
theory to study the gluon and the ghost propagators in a gauge (the so-called
$\lambda$-gauge) interpolating between the Landau and the Coulomb gauge.
Results are compared to the usual minimal Landau gauge.
\end{abstract}

\maketitle

Green's functions encode completely the non-perturbative properties of a quantum
field theory. Thus, the infrared (IR) behavior of the gluon propagator $D(p)$
in Yang-Mills theory should contain information on confinement. In particular,
a violation of reflection positivity (at large separation) for the gluon propagator
is considered an unambiguous signal for the confinement of gluons \cite{Alkofer:2000wg}.
This issue has been widely studied in Landau gauge. Continuum methods, such
as Dyson-Schwinger equations find such a behavior for the gluon propagator \cite{Alkofer:2000wg}.
A clear sign of violation of reflection positivity for the gluon is also obtained
using lattice numerical simulations in the 3d case \cite{Cucchieri:2004mf} while, in 4d,
finite-size effects and technical limitations have not allowed yet
a verification of this result \cite{Cucchieri:2006xi}.
At the same time, the Landau ghost propagator shows a divergent behavior
\cite{Alkofer:2000wg,Cucchieri:2006xi}, stronger than
$p^{-2}$ for small momenta $p$. This strong IR divergence corresponds to
a long-range interaction in real space, which may be related to quark confinement.
On the other hand, in order to understand the confinement mechanism, it is important
to verify how these IR features are modified when considering other
gauges \cite{cucchieri}.
Thus, it is important to investigate these correlation functions in a wide
class of gauges. Particularly important are gauges depending on a parameter,
since one can interpolate between different gauge conditions. 
This allows one to investigate changes in the correlations functions while
varying smoothly the gauge parameter (i.e.\ the gauge condition).

Let us recall that the gauge-fixed Lagrangian density in the (Euclidean)
continuum can be written as
\be
{\cal L} \;= \;\frac{1}{4g^2}F_\mn^a F_\mn^a \,+\, 
          \frac{1}{2\xi}\pdm' A_\mu^a\pd_\nu' A_\nu^a \,+\,
           \bar c^a\pd'_\mu D^{ab}_\mu c^b  \; .
\ee
Here, $\pd_\mu' = \lambda\pd_0 \,+\, \pd_i$,
$\; A_\mu^a$ is the gauge field, $D^{ab}_\mu$ is the usual adjoint covariant derivative,
$F_\mn^a$ is the field-strength tensor
and the ghost fields $\bar c^a$ and $c^a$ have been introduced in order to exponentiate
the Jacobian, i.e.\ the Faddeev-Popov operator $\pd'_\mu D^{ab}_\mu$, introduced
by fixing the gauge. Note that, for $\lambda=1$, the quadratic term in
$\pd'_\mu A_\mu^a$ corresponds to the usual perturbative definition of linear
covariant gauges with gauge parameter $\xi$. In particular, $\xi=1$ is the
Feynman gauge and the limit $\xi=0$ corresponds to Landau gauge. On the other
hand, for $\xi=0$, the parameter $\lambda$ allows to interpolate between
the Landau gauge ($\lambda=1$) and a Coulomb-like gauge ($\lambda=0$). Note that
the usual Coulomb gauge $\pd_i A_i^a = 0$ has to be supplemented by further constraints
\cite{Fischer:2005qe} to yield the limit $\lambda=0$.

Here, we present results for the case $\xi=0$ (the so-called $\lambda$-gauge)
obtained using lattice numerical simulations for the $SU(2)$ gauge group.
To be able to reach the IR asymptotic regime, we consider the 3d case,
extending earlier studies of this subject \cite{Cucchieri:2001tw}.
Clearly, for $\lambda < 1$, Euclidean invariance is explicitly broken by the
gauge condition. In this case, there are two independent kinematic variables,
namely $p_0$ and $|\vec p|$. At the same time, the gluon propagator $D_\mn(p_0,|\vec p|)$
also has two independent tensor structures: its temporal component $D_{00}(p_0,|\vec p|)$
and its spatial projection $D^{\mathrm{tr}}(p_0,|\vec p|)$.
Predictions exist \cite{Fischer:2005qe} that both components of the gluon propagator
vanish at zero total momentum, $p_0^2+|\vec p|^2 = 0$.

\begin{figure}
\includegraphics[width=\textwidth]{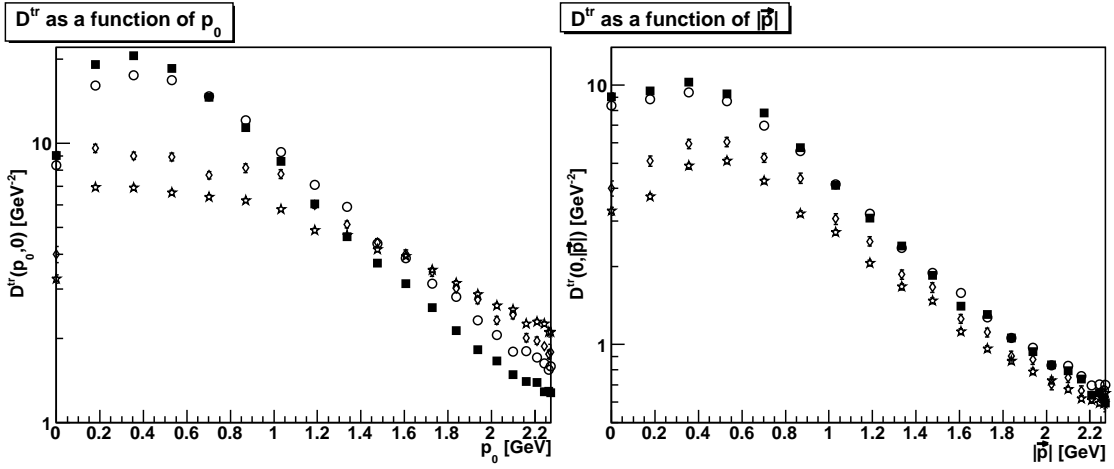}
\caption{The gluon propagator $D^{\mathrm{tr}}(p_0,|\vec p|)$
as a function of $\lambda$ for the lattice volume $40^3$ at $\beta=4.2$.
In the left panel we report data for
$D^{\mathrm{tr}}(p_0,0)$ while in the right panel we consider
$D^{\mathrm{tr}}(0,|\vec p|)$.
We represent $\lambda=1$ with black squares, $\lambda=1/2$ with
open circles, $\lambda=1/10$ with diamonds and $\lambda=1/20$ with stars.}
\label{dtr}
\end{figure}

\begin{figure}
\includegraphics[width=0.5\textwidth]{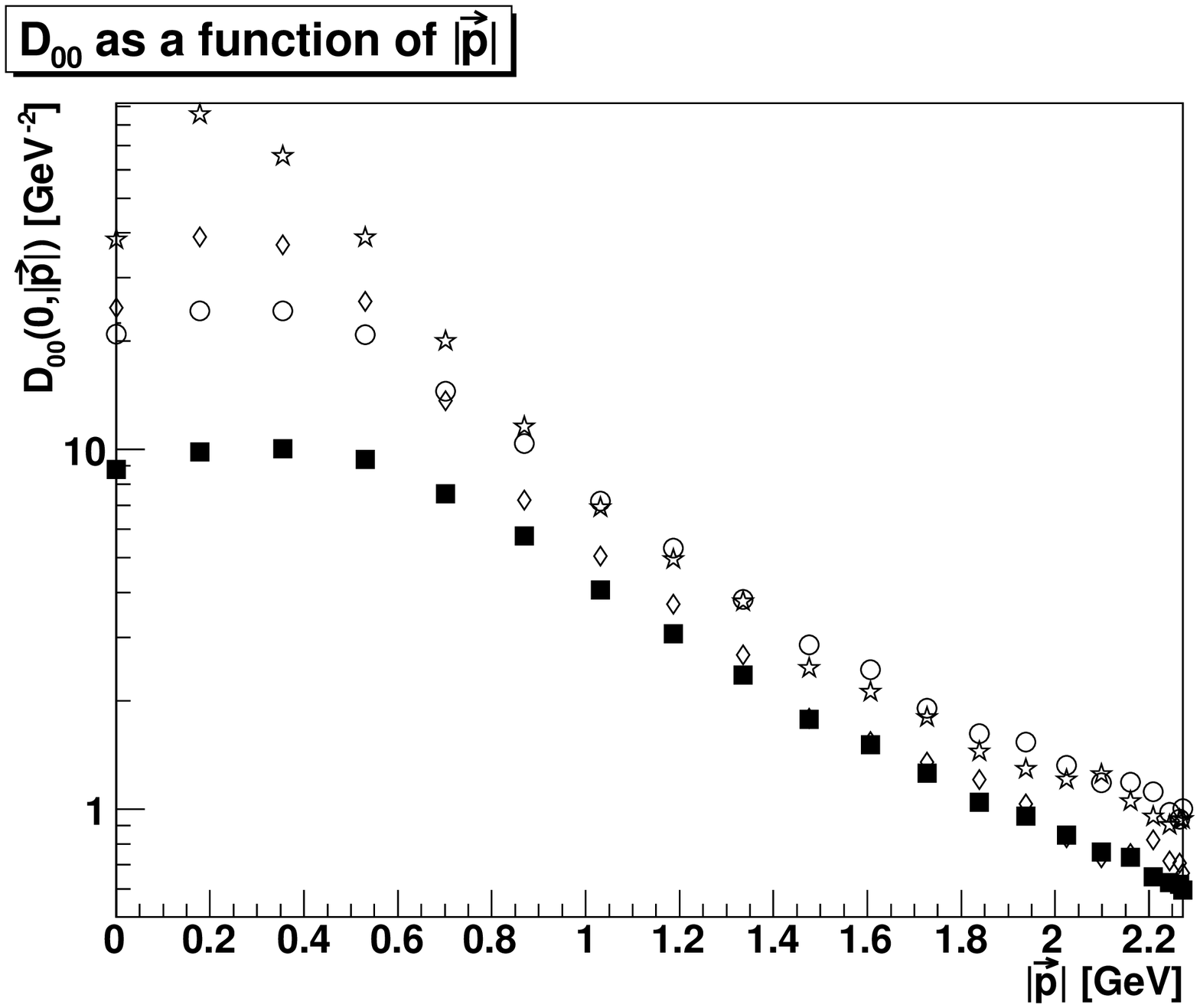}
\includegraphics[width=0.5\textwidth]{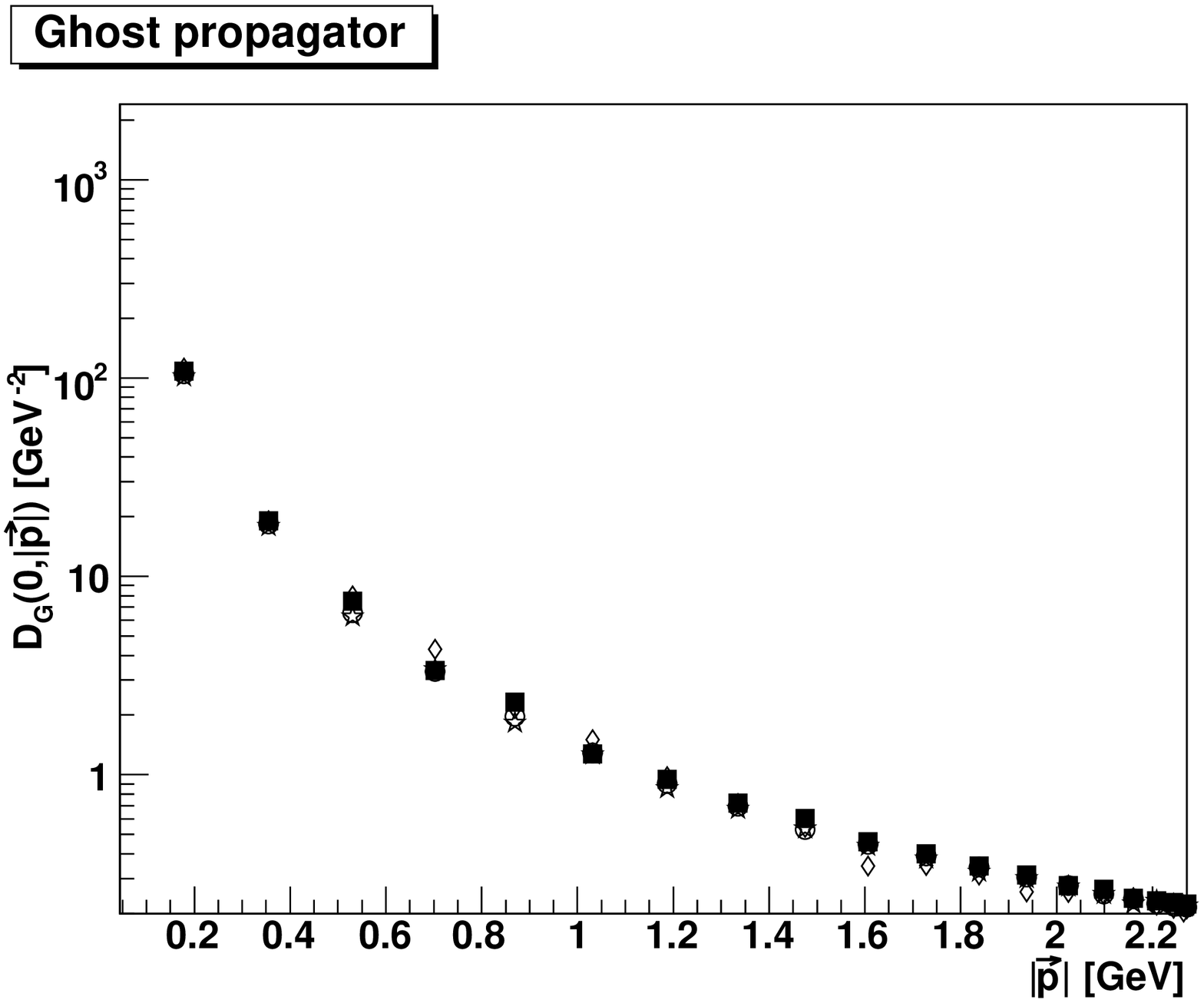}
\caption{In the left panel we show results for the gluon propagator $D_{00}(0,|\vec p|)$,
while in the  right panel we report data for the ghost propagator $D_{G}(0,|\vec p|)$.
In both cases we consider various values of $\lambda$ (symbols and
lattice as in Figure \ref{dtr}).
Note that the quantity $D_{00}(p_0,0)$ vanishes identically due to the gauge condition.
}
\label{oth}
\end{figure}

In Figure \ref{dtr} we report results for $D^{\mathrm{tr}}(p_0,|\vec p|)$ as a separate function
of $p_0$ and of $|\vec p|$ for different values of $\lambda$. We find that, as
a function of $|\vec p|$, the maximum of $D^{\mathrm{tr}}$ moves to larger momenta as
$\lambda$ decreases. On the contrary, when considering its dependence on
$p_0$, we see that the maximum of $D^{\mathrm{tr}}$ moves to smaller momenta and that,
already for $\lambda=1/10$, the maximum can no longer be resolved.
In both cases, for small momenta, the propagator decreases as $\lambda$ gets smaller.
The function $D_{00}(0,|\vec p|)$ is considered in the left panel of Figure \ref{oth}.
In this case, as $\lambda$ decreases, the maximum of the propagator moves
to smaller momenta while the propagator increases in the IR.
This is somewhat expected, since in Coulomb gauge this tensor component diverges
\cite{Cucchieri:2000gu}. Thus, as $\lambda$ decreases, the gluon propagator exhibits
more and more a Coulomb-like behavior.
Finally, the $\lambda$ dependence for the ghost propagator is shown in
the right panel of Figure \ref{oth}. We see that, as a function of $|\vec p|$,
it is essentially independent of $\lambda$.
On the other hand, when $D_G(p_0,|\vec p|)$ is considered as a function of $p_0$
we find that it increases approximately linearly with
$\lambda$ \cite{c2}.
These results support the findings of recent DSE studies \cite{Fischer:2005qe}.
A more
extensive numerical study for this interpolating gauge will be presented elsewhere \cite{c2}.

Finally, let us note that the study of the linear covariant gauges on the lattice
is much more involved. In particular, at $\xi>0$ and $\lambda=1$,
the continuum gauge condition $\pd A_\mu^a(x) = \Lambda^a(x)$
cannot be implemented for any finite $\beta$
because $\Lambda^a(x)$ has a Gaussian distribution, while
the values of $\pd A_\mu^a(x)$ are clearly bounded. Results for the linear covariant
gauges will be presented elsewhere \cite{c3}.

\begin{theacknowledgments}
A.\ M.\ is grateful to the organizers for the opportunity to present this work.
A.\ M.\ was supported by the DFG under grant numbers MA 3935/1-1 and MA 3935/2-1.
A.\ C.\ and T.\ M.\ were supported by FAPESP (under grants \# 00/05047-5
and 05/59919-7) and by CNPq.
\end{theacknowledgments}

\end{document}